\documentclass[fleqn,usenatbib]{mnras}

\usepackage{newtxtext,newtxmath}

\usepackage[T1]{fontenc}

\DeclareRobustCommand{\VAN}[3]{#2}
\let\VANthebibliography\thebibliography
\def\thebibliography{\DeclareRobustCommand{\VAN}[3]{##3}\VANthebibliography}

\usepackage{graphicx}	
\usepackage{amsmath}	
\usepackage{graphicx}	
\usepackage{subcaption} 
\graphicspath{ {./images/} }

\usepackage{amsmath}	
\usepackage{amssymb}    
\usepackage{bm}         
\usepackage{physics}    
\usepackage{anyfontsize}

\title[Flux branching in superconducting NS]{Magnetic coupling through flux branching of adjacent type-I and -II superconductors in a neutron star}

\author[K. H. Thong and A. Melatos]{
K. H. Thong$^{1,2}$
\thanks{E-mail: kokhongt@student.unimelb.edu.au}
and
A. Melatos$^{1,2}$
\thanks{E-mail: amelatos@unimelb.edu.au}
\\
$^{1}$School of Physics, University of Melbourne, Parkville, Victoria, 3010 Australia\\
$^{2}$OzGrav, Australian Research Council Centre of Excellence for Gravitational Wave Discovery, University of Melbourne, Parkville, Victoria, 3010 Australia
}

\date{Accepted XXX. Received YYY; in original form ZZZ}

\pubyear{2024}

\begin{document}
\label{firstpage}
\pagerange{\pageref{firstpage}--\pageref{lastpage}}
\maketitle

\begin{abstract}
The inner and outer cores of neutron stars are believed to contain type-I and -II proton superconductors, respectively. The type-I superconductor exists in an intermediate state, comprising macroscopic flux-free and flux-containing regions, while the type-II superconductor is flux-free, except for microscopic, quantized flux tubes. Here, we show that the inner and outer cores are coupled magnetically, when the macroscopic flux tubes subdivide dendritically into quantized flux tubes, a phenomenon called flux branching. 
An important implication is that up to $\sim 10^{12} (r_1/10^6 \, {\rm cm}) \, {\rm erg}$ of energy are required to separate a quantized flux tube from its progenitor macroscopic flux tube, where $r_1$ is the length of the macroscopic flux tube.  Approximating the normal-superconducting boundary as sharp, we calculate the magnetic coupling energy between a quantized and macroscopic flux tube due to flux branching as a function of, $f_1$, the radius of the type-I inner core divided by the radius of the type-II outer core. Strong coupling delays magnetic field decay in the type-II superconductor. 
For an idealised inner core containing only a type-I proton superconductor and poloidal flux, and in the absence of ambipolar diffusion and diamagnetic screening, the low magnetic moments ($\lesssim 10^{27} \, {\rm G \, cm^3}$) of recycled pulsars imply $f_1 \lesssim 10^{-1.5}$.
\end{abstract}




\section{Introduction}
Neutron stars are believed to contain a type-II proton superconductor, located mainly in the outer core, and a type-I proton superconductor, located mainly in the inner core \citep{sedrakian_type_1997,jones_type_2006,glampedakis_magnetohydrodynamics_2011,haskell_investigating_2013,haber_critical_2017}. The geometry of the interface between the two types of superconductivity is complicated in general \citep{lander_magnetic_2013}. Typical neutron stars have magnetic field strength $B \sim 10^{12} {\rm \ G} \ll H_{\rm c} \sim H_{\rm c1} \sim 10^{15} {\rm \ G}$, where $H_{\rm c}$ is the critical magnetic field of the type-I superconductor and $H_{\rm{c1}}$ is the first critical magnetic field of the type-II superconductor. 
Consequently, the equilibria of both the type-I and -II superconductors are Meissner states, where magnetic flux is expelled ultimately from the bulk. 
However, due to the high electrical conductivity of the relativistic electrons in the core, the flux expulsion time-scale is greater than the nucleation time-scale and the age of the star \citep{baym_superfluidity_1969}, so the proton superconductor stays in a metastable state threaded by normal regions containing magnetic flux. 
The topologies of the normal regions depend on the history of the electrical currents in the star and the surface energy, $\gamma$, of the normal-superconducting boundary. Type-II superconductors have $\gamma < 0$, favoring microscopic normal regions (quantized flux tubes) with magnetic flux quantum $\Phi_0 = \pi \hbar/2e = 2.07 \times 10^{-7} \text{ G cm}^2$. In contrast, type-I superconductors have $\gamma > 0$, favoring macroscopic normal regions with magnetic flux $n\Phi_0$ and $n \gg 1$ \citep{tinkham_introduction_2004}. 

Magnetic field lines cannot terminate at the type-I-II interface because of Gauss' law for magnetism. Hence it is natural to ask how the poloidal fluxes in the type-I and -II superconductors are connected. 
Recent studies have shown that type-1.5 proton superconductivity is expected at the type-I-II interface due to the interpenetrating and strongly interacting neutron superfluid; the neutrons establish a second attractive length-scale for the quantized flux tubes \citep{haber_critical_2017,wood_superconducting_2022}. 
But the puzzle remains: how do the field lines from different phases connect?
For example, near the type-I-II interface, do discrete, quantized flux tubes from the outer core coalesce into smoothly continuous flux in the inner core? 
If the quantized flux tubes coalesce into macroscopic flux tubes near the type-I-II interface but remain separate in the outer core, what energy would be required for a quantized flux tube to wholly separate from the macroscopic flux tube to which it is connected in the inner core and move freely?
In this paper, we explore the possibilities and consequences of flux-tube coalescence via flux branching, where the macroscopic normal regions from the inner core branch out to form quantized flux tubes in the outer core.

Flux branching has been observed terrestrially near the surface of type-I superconductors with $\gamma > 0$ in a uniform, applied magnetic field \citep{huebener_magnetic_2001,han_fractal_2002,tinkham_introduction_2004}. 
Branching in type-I superconductors costs free energy, but the reduced magnetic energy outside the superconductor compensates the cost. 
Bunched-up field lines in the macroscopic normal regions become more uniform spatially near the surface, so that magnetic field lines within the normal regions join more smoothly with those outside \citep{landau_intermediate_1938, landau_theory_1943}. 
There are various ways to branch and much work has been done to find the optimal energy scaling laws for different branching types \citep{landau_theory_1943,andrew_intermediate_1948,hubert_zur_1967,choksi_energy_2004,conti_branched_2015}. We employ the flux branching formalism developed by \citet{landau_theory_1943} to consider the possibility of branching near the type-I-II interface in neutron stars.

A striking similarity exists between the boundary of a type-I superconductor in a uniform, applied magnetic field and the type-I-II interface. 
Bunched-up field lines are preferred inside the type-I superconductor, while more uniform field lines are preferred outside the type-I superconductor and inside the type-II superconductor. 
Drawing intuition from flux branching in type-I superconductors immersed in a uniform, applied field, one expects flux branching to occur also at the interface of a type-I and -II superconductor.
Conversely, if flux branching does not occur, are the macroscopic normal regions and quantized flux tubes separately closed, so that the type-I and -II superconductors are magnetically disconnected \citep{kozlov_closed_1993,lander_magnetic_2013,anzuini_differential_2020,cadorim_closed_2023}? Or is the whole core filled with mesoscopic normal regions? Each possibility mentioned above has different implications for observable phenomena in a neutron star, e.g., secular spin down \citep{beskin_physics_1993,gurevich_energy_2007,barsukov_long-term_2014}, rotational glitches \citep{link_constraining_2003,haskell_investigating_2013,drummond_stability_2018,sourie_vortex_2020,thong_stability_2023}, magnetic evolution, e.g., the time-scale of magnetic decay \citep{srinivasan_novel_1990} and polar magnetic drift \citep{macy_pulsar_1974,shaham_free_1977,alpar_neutron_1987,melatos_radiative_2000}, and gravitational wave emission \citep{melatos_persistent_2015}.

The paper is structured as follows. In Section \ref{sec:2}, we review the canonical flux branching formalism introduced by \citet{landau_theory_1943} and apply energetic arguments to flux branching in a neutron star. 
We find that flux branching is the preferred mechanism to magnetically couple the type-I and -II superconductors under a wide range of plausible parameters.
In Section \ref{sec:3}, we calculate the coupling energy between a macroscopic and quantized flux tube of the same flux tree.
We then discuss an important astrophysical implication of flux branching in Section \ref{sec:4}, namely how magnetic coupling retards magnetic decay, lengthening the time-scale over which flux remains ``frozen'' in the inner and outer cores. The observed magnetic moments of recycled pulsars are used to set an upper bound on the volume of the type-I superconductor based on the above phenomenon. 


\section{Free Energy of Flux Branching}
\label{sec:2}

In this section, we apply the sharp-interface model introduced by \citet{landau_theory_1943} and later developed by others \citep{andrew_intermediate_1948,choksi_energy_2004,choksi_ground_2008,conti_branched_2015,cinti_interpolation_2016,conti_branched_2018} to study the behaviour of flux branching in a neutron star. 
The sharp-interface model is outlined in Section \ref{sec:2.1}. 
We calculate the free energy of a single branching cell in Section \ref{sec:2.2} and concatenated branching cells or a flux tree in Section \ref{sec:2.3}. 
The expected size of the macroscopic normal regions is briefly discussed in Section \ref{sec:2.4}. 
Applying the results to neutron stars, we show in Section \ref{sec:2.5} that the free energy of flux branching is negative under a range of plausible conditions, implying that flux branching occurs.

\subsection{Sharp-interface model}
\label{sec:2.1}
The sharp-interface model can be derived from the Ginzburg-Landau model in the limit that the coherence length, $\xi$, and the London penetration depth, $\lambda$, are much less than the size of the normal and superconducting regions \citep{chapman_asymptotic_1995,choksi_energy_2004}. 
In the sharp-interface model, normal regions have a uniform magnetic field $B = H_{\rm c}$, while superconducting regions have $B = 0$. The surplus free energy, $F$, when including a normal region in an otherwise superconducting material, is given by
\begin{equation}
\label{eq:F1}
    F = \frac{H_{\rm c}^2V}{4\pi} + \frac{H_{\rm c}^2 }{8\pi} \int dS \, \delta
\end{equation}
in cgs units, where $V$ is the volume of the normal region and $S$ and $\delta$ are the surface area and the scaled surface energy of the normal-superconducting boundary, with $\delta = {8\pi} {H_{\rm c}}^{-2}\gamma \approx \xi - \lambda$ \citep{tinkham_introduction_2004}. Type-I superconductors have $\delta > 0$, while type-II superconductors have $\delta < 0$. For a given magnetic flux, $\Phi$, type-I superconductors energetically favor $\Phi$ to be contained within a macroscopic normal region. In contrast, type-II superconductors energetically favor $\Phi$ to be spread across many microscopic normal regions, often called quantized flux tubes (see Appendix \ref{appendix:A}). 
As a note of caution, quantized flux tubes have magnetized cores of size $\sim \lambda$ and normal regions of size $\sim \xi$, so the sharp-interface approximation applies poorly to quantized flux tubes.
Indeed, normal regions are modelled as non-interacting in the sharp-interface approximation, whereas quantized flux tubes are mutually and magnetically repulsive. We briefly discuss some improvements to the sharp-interface model in this direction in section \ref{sec:2.5} and leave a more detailed analysis for a future paper.

\subsection{Single cell}
\label{sec:2.2}
\begin{figure}
    \centering
    \includegraphics[width=0.5\textwidth]{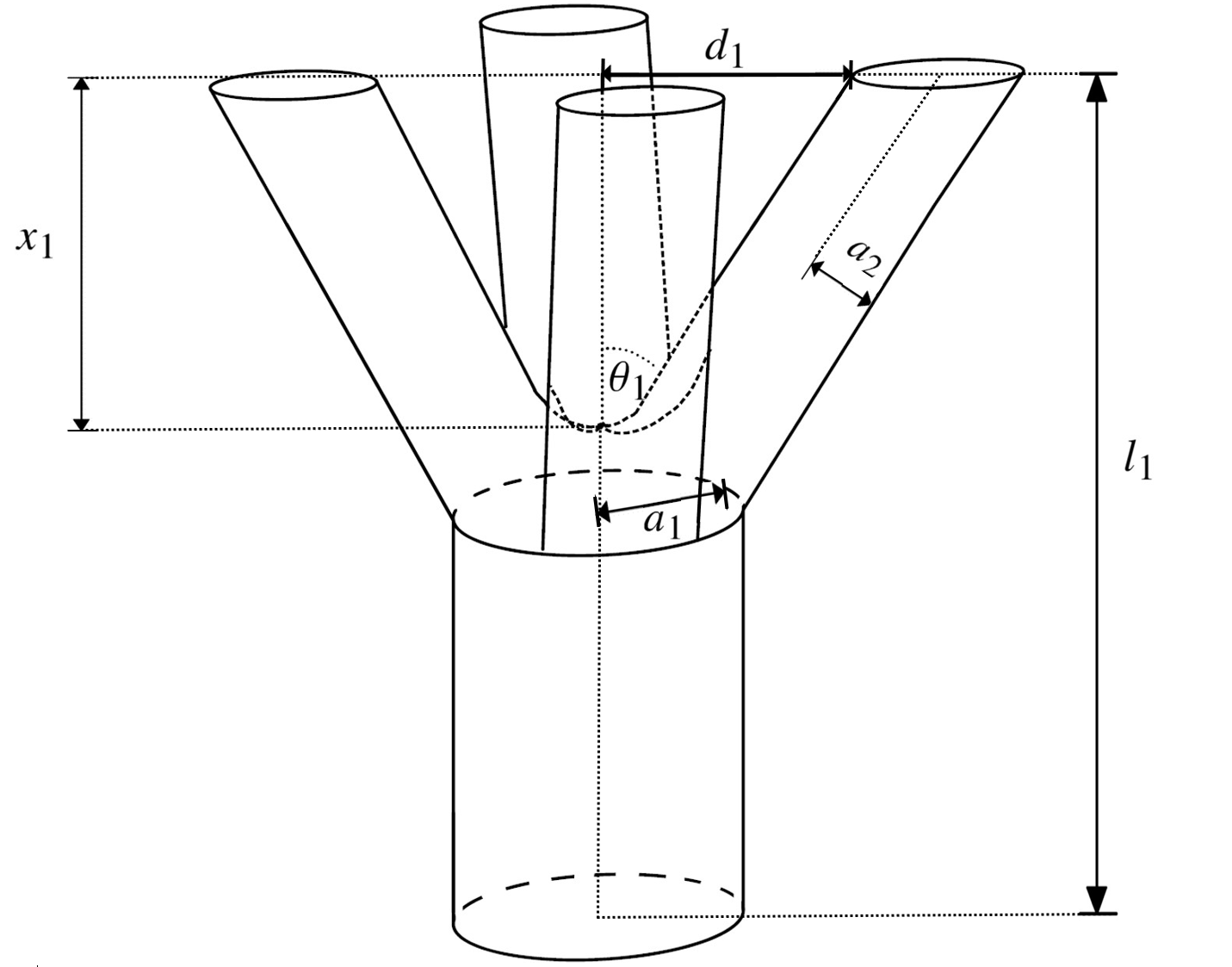}
    \caption{Schematic of a branching cell adapted from \citet{andrew_intermediate_1948}, where a macroscopic normal region (bottom cylinder) branches into four smaller normal regions (top cylinders).}
    \label{fig:1}
\end{figure}
We follow \citet{andrew_intermediate_1948} and consider the branching of a macroscopic normal region (the trunk) into four smaller normal regions (the branches). The trunk and branches collectively form a cell, which is sketched schematically in Figure \ref{fig:1}. 
Other types of branching are also possible \citep{landau_theory_1943,andrew_intermediate_1948}, but we do not investigate them in this paper. Flux conservation means that the total cross-sectional area of the normal regions is conserved, viz
\begin{equation}
\label{eq:flux_conservation}
    \pi a_1^2 = 4\pi a_2^2,
\end{equation} 
where $a_1$ and $a_2$ denote the cylindrical radius of the trunk and any of the four branches, respectively. (Recall that we have $B = H_{\rm c}$ uniformly in a normal region.) 
Assuming that the branches extend nearly parallel to the symmetry axis of the trunk (i.e.\ $\theta_1 \ll 1$, where $\theta_1$ is the branch half-opening angle labelled in Figure \ref{fig:1}) as in a neutron star, the curved surface area, $S_1$, and the volume, $V_1$, of the branching cell are given by
\begin{align}
    S_1 = 2\pi a_1 \left(l_1 + x_1 + \frac{d_1^2}{x_1}\right), 
    \label{eq:S2}
\end{align}
and 
\begin{align}
    V_1 = \pi a_1^2 \left(l_1 + \frac{d_1^2}{2x_1} \right),    
\end{align}
respectively, where $l_1$ is the vertical height of the whole cell, $x_1$ is the vertical height of the branches, and we have $d_1 = x_1 \tan\theta_1$ (half the horizontal distance between the axis of the trunk and a branch), as in Figure \ref{fig:1}.
Hence, from equation (\ref{eq:F1}), the surplus free energy, $F_1$, of a branching cell is 
\begin{equation}
    F_1 = \frac{H_{\rm c}^2}{4\pi} \pi a_1^2 \left(l_1 + \frac{d_1^2}{2x_1} \right) + \frac{H_{\rm c}^2\delta_1}{8\pi} 2\pi a_1 \left(l_1 +x_1 + \frac{d_1^2}{x_1}\right),
\end{equation}
where $\delta_1$ refers to $\delta$ in the specific cell.

\subsection{Multiple cells}
\label{sec:2.3}
\begin{figure}
    \centering
    \includegraphics[width=0.4\textwidth]{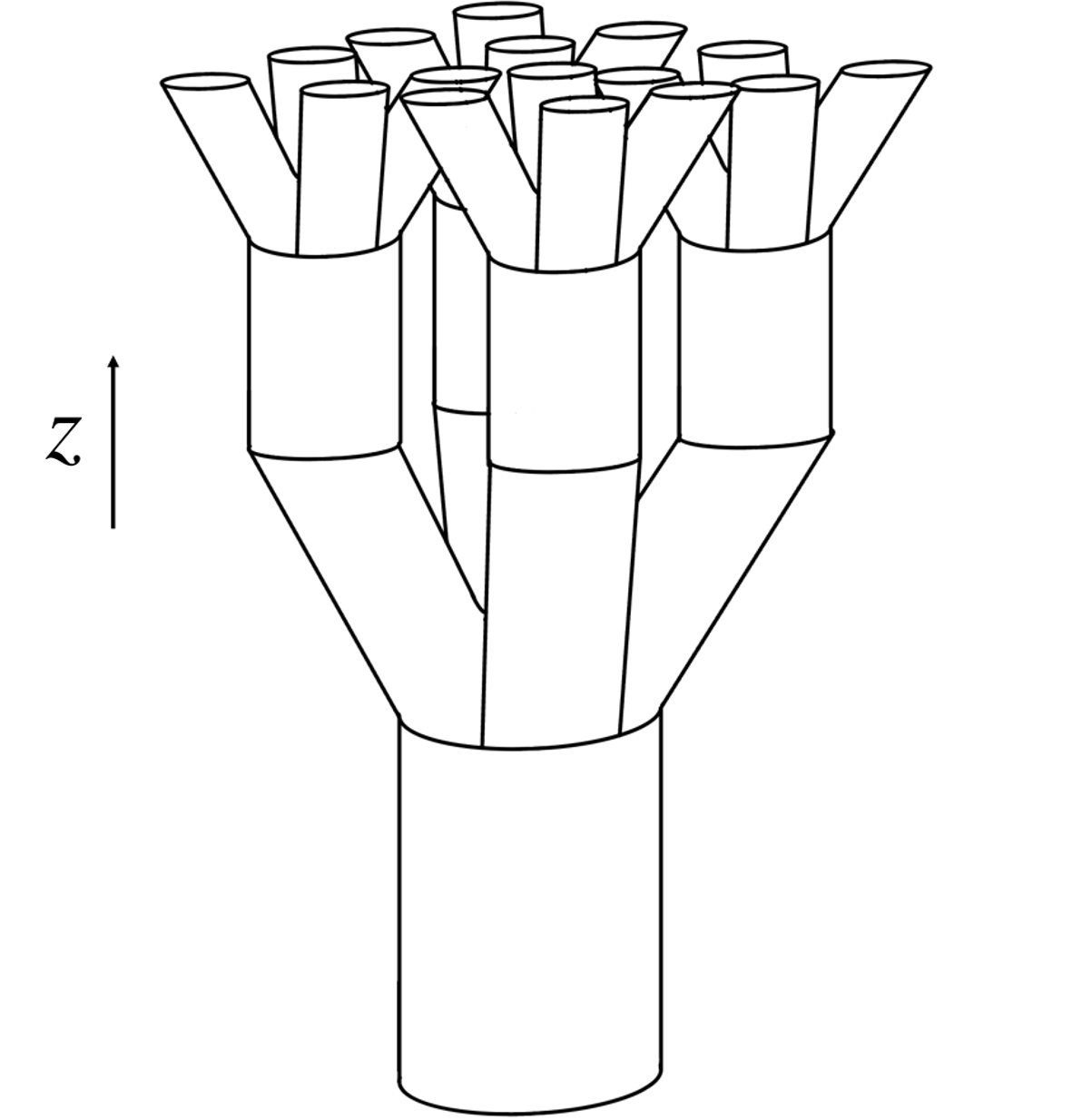}
    \caption{Schematic of two stacked branching levels containing five branching cells. On the lower level, a macroscopic normal region branches into four smaller normal regions. On the upper level, the four smaller normal regions branch into sixteen even smaller normal regions.}
    \label{fig:2}
\end{figure}
What is the total surplus free energy of a flux tree with an arbitrary number of branching cells? 
By way of illustration, consider $(4^{N-1}-1)/3$ copies of the arrangement in Figure \ref{fig:1}, concatenated across $N-1$ branching levels. 
In Figure \ref{fig:2}, for example, five branching cells are stacked over two levels to form a flux tree. 
We employ cylindrical coordinates to describe the flux tree. 
Let the $z$-axis be the branching axis, i.e.\ the axis along which $\delta(z)$ varies, so that $-L_1/2 \leq z < 0$ contains the type-I superconductor, and $0 < z \leq L_2/2$ contains the type-II superconductor. 
The $z$-axis is parallel to the trunks of the branching cells.

To find the total surplus free energy of a flux tree, we must first find $\delta(z)$.
We have $\delta(z) \approx \xi - \lambda$ qualitatively \citep{tinkham_introduction_2004}, where $\xi$ and $\lambda$ are functions of the proton density \citep{mendell_magnetohydrodynamics_1998}.
Given an equation of state of a neutron star, $\delta(z)$ can be calculated numerically within the Ginzburg-Landau framework; refer to \citet{tinkham_introduction_2004} for details. 
However, an exact calculation of $\delta(z)$ is complicated and lies outside the scope of the paper.
In this paper, for simplicity, we assume that $\delta(z)$ is a linear function of $z$ in the core of an idealised neutron star, viz.
\begin{equation}
    \delta(z) = \frac{2\delta_0z}{L_2},
\end{equation}
where $-\delta_0 \sim 10^{-12} \rm{\, cm}$ is the scaled surface energy at $z = L_2/2$. 
We make a further approximation, and discretise $\delta(z)$ according to
\begin{equation}
\label{eq:discreteapprox}
    \delta_n = \frac{2\delta_0z_n}{L_2},
\end{equation}
where $\delta_n$ is the uniform value of $\delta(z)$ in the $n$-th branching cell, and $z_n$ is the $z$-coordinate of the base of the $n$-th branching cell.
Flux conservation implies
\begin{equation}
\label{eq:a_n}
    a_n = \frac{a_1}{2^{n-1}},
\end{equation} 
where $a_n$ is the cross-sectional area of a branch at the $n$-th level, and $n=1$ is the bottom level. Applying (\ref{eq:F1}), the total surplus free energy of the flux tree (summed over $N-1$ branching levels) is
\begin{align}
\label{eq:F2}
    F = \frac{H_{\rm c}^2}{4\pi} \pi a_1 \sum_{n=1}^{N-1} \left[a_1 \left(l_n + \frac{d_n^2}{2x_n}\right) + 2^{n-1} \delta_n \left(l_n + x_n + \frac{d_n^2}{x_n} \right) \right].
\end{align}
\begin{figure}
    \centering
    \includegraphics[width=0.4\textwidth]{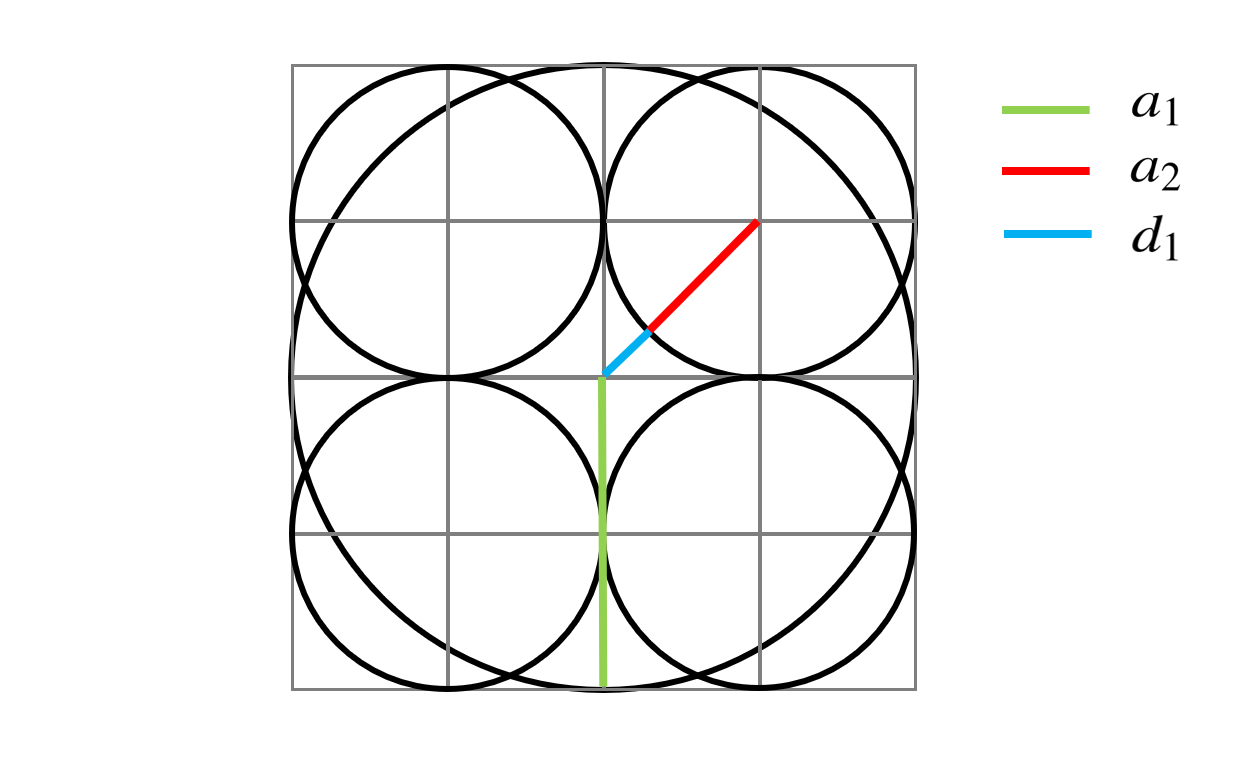}
    \caption{Top view of Figure \ref{fig:1}, but for the minimum allowed $d_1$ before the normal regions overlap. The gray lines serve as a grid, while the black curves mark the normal-superconducting boundaries. The green, red and blue line segments indicate $a_1, \, a_2, \, d_1$, respectively.}
    \label{fig:3}
\end{figure}

Figure \ref{fig:3} shows the top view of the branching cell in Figure \ref{fig:1} with $d_1$ as small as possible. To prevent the normal regions from overlapping, we require $d_1 > \sum_{n=2}^{N-1} d_n$ and $d_1 > (\sqrt{2}-1)a_1/2$, as seen in Figure \ref{fig:3}. We follow \citet{landau_theory_1943} and \citet{andrew_intermediate_1948} and choose
\begin{equation}
\label{eq:d_n}
    d_n = \frac{d_1}{2^{n-1}}.
\end{equation}


\subsection{Size of the macroscopic normal regions}
\label{sec:2.4}
Before calculating $F$, we discuss the expected size of the macroscopic normal regions in a neutron star, i.e. the trunks in Figures \ref{fig:1}--\ref{fig:3}. 
The size of the largest trunk, $a_1$, is determined by various complicated factors including but not limited to the initial conditions of the magnetic field, as well as the density and temperature of the protons and the functional form of $\delta(z)$ \citep{tinkham_introduction_2004}. 
This is a complicated topic without experimental information in the neutron star context. Only order of magnitude estimates exist in the current literature, which postulate $10^{-3} \lesssim a_1/(1 \, {\rm cm}) \lesssim 10^{-1}$ \citep{buckley_neutron_2004,sedrakian_type-i_2005}. 
However, these estimates should be treated cautiously, as $a_1$ probably exhibits vestiges of the trunk's formation history due to the high electrical conductivity in the core. 
For example, suppose an eddy in a convective dynamo of size $\gtrsim 10^{-2} \, {\rm cm}$ and magnetic field $\sim 10^{15} \, {\rm G}$ existed early in the star's life, before the superconducting phases nucleate. 
In that case, a macroscopic flux tube of a similar size is formed and frozen into the type-I superconductor directly, after the superconducting phases nucleate. 
Indeed, a statistical distribution of $a_1$ values is likely, whose properties depend on the properties of the dynamo.
Clarification of these issues requires large-scale magnetohydrodynamic simulations, which fall outside the scope of this paper.
Instead, we study here a representative range of special cases, viz.\ $N = 28, \, 38, \, 48$, corresponding to $a_1 \approx 10^{-3}, \, 1, \, 10^{3}\, {\rm cm}$, respectively, with $a_1 = 2^{N-1}a_{N}$ and $a_{N} = \lambda = 6.7 \times 10^{-12} \, {\rm cm}$ \citep{baym_electrical_1969}.
$N=48$ corresponds to $\approx 2 \times 10^{28}$ quantized flux tubes per flux tree.
Note that there are typically $\sim 10^{30}$ quantized flux tubes in total in the type-II superconductor of the star, implying $N \lesssim 51$.

\subsection{Branching in type-I superconductivity in a neutron star}
\label{sec:2.5}
In this paper, we study the conditions under which flux branching occurs in the type-I superconductor in a neutron star. 
Specifically, we seek to identify the conditions under which the free energy turns negative. From (\ref{eq:F2}), the surplus free energy, $F_{\rm I}$, of a trunk branching into $4^{N-1}$ branches in the type-I region is given by
\begin{align}
    F_{\rm I} = \ &\frac{H_{\rm c}^2}{4\pi} \pi a_1 \sum_{n=1}^{N-1} \Biggl[a_1 \Biggl(l_n + \frac{d_1^2}{2^{2n-1} x_n}\Biggr) \nonumber \\ &+ 2^{n-1} \delta_n \Biggl(l_n + x_n + \frac{d_1^2}{2^{2n-2} x_n} \Biggr) \Biggr],  
\end{align}
with $\sum_{n=1}^N l_n = L_1/2$ and $x_n \leq l_n$. From (\ref{eq:F1}), the surplus free energy, $F_{\rm II}$, of the $4^{N-1}$ terminating rectilinear microscopic normal regions (i.e. quantized flux tubes) in the type-II region is given by
\begin{equation}
    F_{\rm II} = \frac{H_{\rm c}^2}{4\pi} 4^{N-1}\pi a_{N} \left[\frac{a_{N} L_2}{2} +\int_{0}^{L_2/2} dz \, \delta(z) \right].
\end{equation}

The total surplus free energy, $F_{\rm b}$, of the flux tree in the type-I region concatenated with the $4^{N-1}$ branches in the type-II region is 
\begin{equation}
\label{eq:FNI}
    F_{\rm b} = F_{\rm I} + F_{\rm II}.
\end{equation}


We can minimize equation (\ref{eq:FNI}) with respect to $l_n$ and $x_n$, subject to the constraints for $l_n$ and $x_n$ above. 
However, unlike in previous works \citep{landau_theory_1943,andrew_intermediate_1948,conti_branched_2015} where $\delta$ is a constant, the form of $\delta_n$ here depends on
\begin{equation}
    z_n = -\frac{L_1}{2} + \sum_{m=1}^{n-1} l_m, 
\end{equation}
which complicates the minimization.
As a first pass, we do not minimize (\ref{eq:FNI}). 
Instead, we guess a particular choice of $l_n$ and $x_n$ and calculate $F_{\rm b}$. 
We then compare $F_{\rm b}$ with the free energy without flux branching, $F_{\rm nb}$, which we specify below.
We expect branching to be more energetically favorable in regions with lower $\delta(z)$, so we choose 
\begin{equation}
\label{eq:ln}
    l_n = \frac{1}{1-2^{-N}} \frac{L_1}{2^{n+1}},
\end{equation}
to ensure that most branching occurs near the type-I-II interface.
For simplicity, we choose $x_n = l_n/2$.
In the regime $4^{N-1} \gg 1$, as in neutron stars, equation (\ref{eq:FNI}) evaluates to 
\begin{align}
\label{eq:FNI_i}
    F_{\rm b} \approx \ &\frac{H_{\rm c}^2a_1}{4} \Biggl[2^{N-3}\delta_0 L_2  + a_1 \left(\frac{L_1+L_2}{2} \right)   + \frac{8a_1 d_1^2}{L_1} \nonumber \\
    &-\frac{\delta_0}{L_2}\Biggl(\frac{L_1^2}{2}+\frac{32 d_1^2}{3} \Biggr) \Biggr].
\end{align}

In this paper, we assume that the flux trees are as wide as they can be without overlapping, yielding 
\begin{equation}
\label{eq:d1}
    d_1 = \sqrt{\frac{\pi (L_1/2)^2 }{8 N_{\rm ft}}} - a_2,
\end{equation}
where $N_{\rm ft}$ is the number of flux trees implied by flux conservation, given by 
\begin{equation}
\label{eq:Nft}
    N_{\rm ft} = \frac{B (L_1/2)^2}{H_{\rm c} a_1^2}.
\end{equation}
Equations (\ref{eq:d1}) and (\ref{eq:Nft}) are justified in Appendix \ref{appendix:B}.



Is flux branching favored over other plausible configurations? 
Suppose the magnetic topology between the type-I and -II superconductors is open, i.e.\ every field line in the type-I region also threads the type-II region. 
In that case, one plausible alternative without flux branching is where the entire type-I and -II superconductors are filled with $m$ rectilinear normal regions\footnote{Recall that the flux expulsion time-scale is greater than the nucleation time-scale and the age of the star \citep{baym_electrical_1969}, so magnetic flux threads both the type-I and -II superconductors. 
The true ground state of the superconductors in neutron stars is the Meissner state, where both type-I and -II superconductors are flux-free.} of a fixed radius, $a_1/\sqrt{m}$ (from flux conservation), and length $(L_1+L_2)/2$, with $1 \leq m \leq 4^{N-1} $. 
From (\ref{eq:F1}), the free energy without flux branching, $F_{\rm nb}$, is given by
\begin{equation}
    F_{\rm nb} = \frac{H_{\rm c}^2}{4\pi} m\pi \frac{a_1}{\sqrt{m}} \left[\frac{a_1}{\sqrt{m}}\left(\frac{L_1+L_2}{2} \right)+\int_{-L_1/2}^{L_2/2} dz \, \delta(z)\right],
\end{equation}
which evaluates to
\begin{equation}
\label{eq:FNno}
    F_{\rm nb} = \frac{H_{\rm c}^2 a_1}{4} \left[a_1 \left(\frac{L_1+L_2}{2} \right) + \frac{\sqrt{m}\delta_0  \left(L_2^2-L_1^2\right)}{4 L_2}  \right].
\end{equation}
We seek to compare $F_{\rm b}$ with $F_{\rm nb}$ for anything between $m=1$ macroscopic flux tube to $m=4^{N-1}$ quantized flux tubes.
Comparing (\ref{eq:FNI_i}) with (\ref{eq:FNno}), the excess free energy of flux branching, $\Delta F = F_{\rm b} - F_{\rm nb}$, is given by
\begin{align}
\label{eq:DF}
    \Delta F = \ &\frac{H_{\rm c}^2a_1}{4} \Biggl[\left(\frac{2^{N-1}-\sqrt{m}}{4}\right)\delta_0 L_2   + d_1^2 \Biggl( \frac{8a_1}{L_1} - \frac{32 \delta_0}{3L_2}\Biggr) \nonumber \\ &+\frac{(\sqrt{m}-2)\delta_0  L_1^2}{4 L_2} \Biggr].
\end{align}
If, instead, the magnetic topologies of the type-I and -II superconductors are separately closed, the situation is more complicated. One must study ring-shaped macroscopic regions and quantized flux tubes and their stabilities \citep{kozlov_closed_1993,anzuini_differential_2020,cadorim_closed_2023}, which lie outside the scope of this paper.


To gain some intuition from (\ref{eq:DF}), we first consider a special case where the type-I and -II superconductors fill the entire core, with $L_1 = L_2 \sim 10^6 \, {\rm cm}$. In this example, (\ref{eq:DF}) simplifies to
\begin{equation}
\label{eq:DFspecial}
    \Delta F \approx \frac{H_{\rm c}^2a_1}{4} \Biggl[\frac{2^{N-1}\delta_0 L_2}{8}   + \frac{8a_1 d_1^2}{L_1}  \Biggr]. 
\end{equation}
For the three representative examples $(N, a_1) =  (28, \, 10^{-3}\, {\rm cm}), \, (38, \, 1\, {\rm cm}), \, (48,\, 10^{3}\, {\rm cm})$, $d_1 = 19 a_1$ and $\delta_0 = -10^{-12} \, {\rm cm}$, we obtain $\Delta F < 0$, such that branching is favored.
We find that the magnitude of the first term greatly exceeds the magnitude of the second term in (\ref{eq:DFspecial}).
More generally, for all $m$ and $(L_1 + L_2)/2 \sim 10^6 \, {\rm cm}$, we obtain $\Delta F < 0$ for $8.6 \times 10^{-5} \lesssim L_2/L_1 \lesssim 1.4 \times 10^4$, $2.7 \times 10^{-6} \lesssim L_2/L_1 \lesssim 1.4 \times 10^2$, and $8.4 \times 10^{-8} \lesssim L_2/L_1 \lesssim 0.87$ for $N = 28, 38,$ and $48$, respectively. 
The viable $L_2/L_1$ range for branching shrinks, as $N$ increases.
For $L_2/L_1 \gg 1$, the preferred state is where $4^{N-1}$ quantized flux tubes thread both superconductors $(m = 4^{N-1})$.
For $L_2/L_1 \ll 1$ the preferred state is where one macroscopic flux tube with radius $a_1$ threads both superconductors ($m=1$).
Otherwise, as long as the sharp-interface approximation holds, branching is favored as it allows a macroscopic flux tube to thread part of the type-I superconductor and $4^{N-1}$ microscopic quantized flux tubes to thread the type-II superconductor.

\section{Flux coupling}
\label{sec:3}
\begin{figure}
    \centering
    \includegraphics[width=0.45\textwidth]{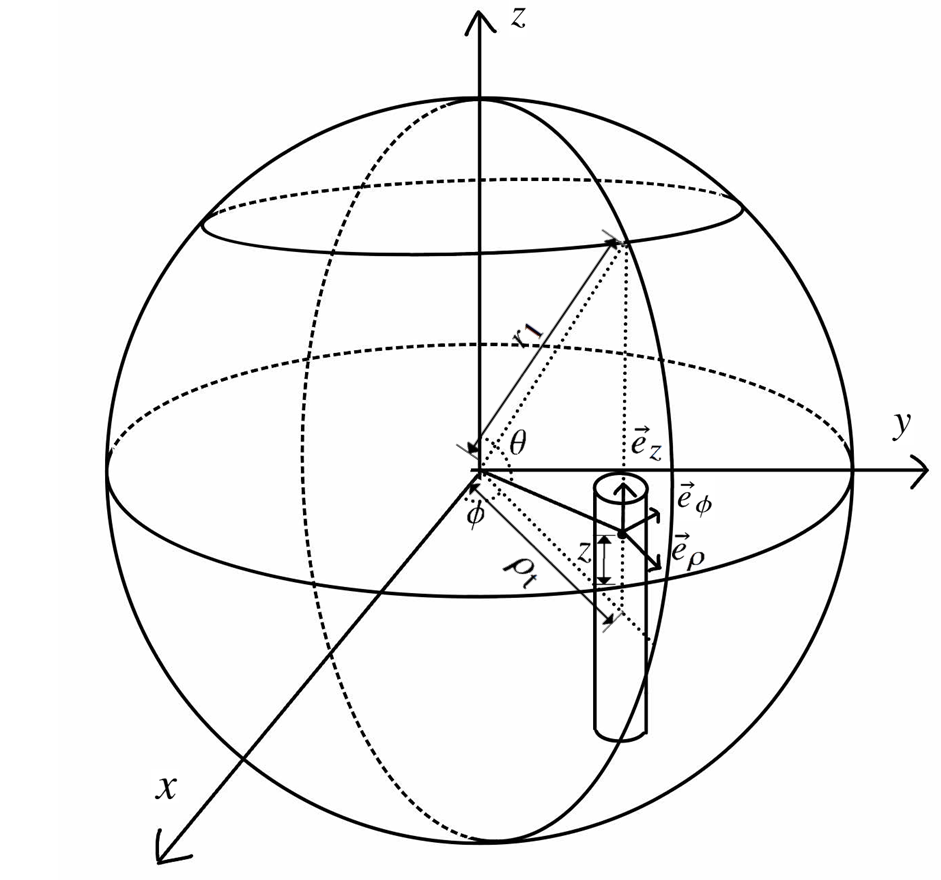}
    \caption{An illustration of the trunk (macroscopic flux tube) of a flux tree at $\rho = \rho_{\rm t}$ in the inner core with $r \leq r_1$. The $z-$axis is the branching axis, while $\delta$ varies as a function of $r$.}
    \label{fig:4}
\end{figure}
In this section, we explore one important consequence of flux branching, namely the coupling of quantized flux tubes in the type-II superconductor with macroscopic flux tubes in the type-I superconductor. 
Such coupling affects the magnetic evolution of the star.
Flux branching couples quantized flux tubes (in the type-II superconductor) with a macroscopic flux tube (in the type-I superconductor) of the same flux tree. 
For a quantized flux tube to move freely in the type-II superconductor, the flux tube has to either (i) separate from the flux tree, or (ii) drag the entire flux tree with $4^{N-1}-1$ other flux tubes along with it.
Option (i) increases $S$ in the type-I superconductor and costs energy, because type-I superconductors have $\delta > 0$.
We define the coupling energy, $E_{\rm c}$, as the energy required to separate a quantized flux tube from the flux tree, as in option (i).
Option (ii), on the other hand, is less probable than option (i). 
A quantized flux tube is unlikely to drag $4^{N-1}-1 \gg 1$ other flux tubes along with it, as other flux tubes may be pinned to inhomogeneities such as neutron superfluid vortices \citep{muslimov_vortex_1985,sauls_superfluidity_1989,srinivasan_novel_1990}.




Let us calculate the energy, $E_{\rm c}$, required to separate a quantized flux tube from the flux tree in the type-I superconductor, as in option (i). 
Specifically, let us calculate $E_{\rm c}$ as a function of the flux tree's location and the type-I superconductor's size.
Suppose the inner core of the star is a type-I superconductor containing flux trees, and the outer core is a type-II superconductor containing quantized flux tubes concatenated with the flux trees, as in Section \ref{sec:2}.
Define spherical $(r, \theta, \phi)$ and cylindrical ($\rho, \phi, z$) coordinates, where the unit vector $\vec{e}_z$ is parallel to the branching axis and the magnetic dipole moment, $\vec{\mu}$, of the star, and $\vec{e}_r$ is the radial unit vector along which $\delta(r) = 2\delta_0 (r-r_1)/r_2$ varies\footnote{In Section \ref{sec:2}, the branching axis and the axis along which $\delta$ varies are the same. Here, we consider a more general scenario where the two axes are not necessarily the same, using spherical and cylindrical coordinates. In these coordinates, we see that the analysis performed in Section \ref{sec:2} is a special case where the flux tree axis is located at $\rho = 0$.}.
Here $0 \leq r \leq r_1$ and $r_1 \leq r \leq r_2$ are the inner (type-I) and outer (type-II) cores of the star, respectively.

Consider the following separation scenario: the macroscopic flux tube (trunk) of a flux tree at $\rho = \rho_{\rm t}$, which we illustrate in Figure \ref{fig:4}, with radius $a_1$ and an arbitrary length $(r_1^2 - \rho_{\rm t}^2)^{1/2}/2$, separates into a smaller trunk with radius $a'_1 < a_1$ and a quantized flux tube with radius $a_1/2^{N-1}$, with $a_1/2^{N-1} \ll a_1 \approx a'_1$.
Flux conservation implies
\begin{equation}
    a'_1 = a_1\sqrt{1-4^{1-N}}
\end{equation}
and $V$ remains unchanged. The change in $S$ per unit length of the flux tree is given by
\begin{equation}
\label{eq:deltas}
    \Delta s \approx 2 \pi a_1(2^{1-N} - 2^{1-2N}),
\end{equation}
implying $\Delta s > 0$.
Applying (\ref{eq:F1}) to the separation scenario, we obtain
\begin{align}
\label{eq:E_c}
    E_{\rm c} &= \frac{H_{\rm c}^2 \Delta s}{4\pi} \int_{0}^{\frac{1}{2}\sqrt{r_1^2-\rho_{\rm t}^2}} dz\,  \delta\left(\abs{z \, x\, y\, \vec{e}_z+\rho_{\rm t}\vec{e}_\rho}\right) \\
\label{eq:E_c2}
    &= \frac{H_{\rm c}^2 \lambda \delta_0 r_2 f_1^2\Lambda(f_\rho)}{2},
\end{align}
where we write $f_1 = r_1/r_2$, $f_\rho = \rho_{\rm t}/r_1$, $\rho_{\rm t}\vec{e}_\rho$ points from the centre of the star to the centre of the intersection between the flux tree and the equatorial plane, and we define
\begin{align}
    \Lambda(f_\rho) = &-\sqrt{1-f_\rho^2}+\frac{1}{4}\sqrt{1+2f_\rho^2-3f_\rho^4} 
    +f_\rho^2\tanh^{-1}{\sqrt{\frac{1-f_\rho^2}{1+3f_\rho^2}}},
\end{align}
with $0 \leq \Lambda(f_\rho) < 1$.

The coupling energy spans the range $0\lesssim E_{\rm c} / (1\, {\rm erg}) \lesssim 10^{13}$ for typical neutron star parameters, depending on the geometry and relative size of the type-I and -II regions. For example, if the type-I and -II regions are of the same size (i.e., $f_1 = 1/2$) and one has $f_\rho \ll 1$, equation (\ref{eq:E_c2}) reduces to
\begin{align}
    E_{\rm c} = \ &6.3 \times 10^{11} \left(\frac{H_{\rm c}}{10^{15} \, \rm{G}}\right)^2\left(\frac{\delta_0}{-10^{-12} \, \rm{cm}}\right) \nonumber \\
    &\times \left(\frac{\lambda}{6.7 \times 10^{-12} \, \rm{cm}}\right)  \left(\frac{r_2}{10^6 \, \rm{cm}}\right) \, {\rm erg}.
\end{align}
If the type-I region is small (i.e., $f_1 \rightarrow 0$), then one has 
\begin{align}
\label{eq:E_cf10}
    E_{\rm c} = \ &3.3 \times 10^{12} \Lambda(f_\rho) f_1^2 \left(\frac{H_{\rm c}}{10^{15}\, \rm{G}}\right)^2\left(\frac{\delta_0}{-10^{-12} \, \rm{cm}}\right) \nonumber \\
    &\times \left(\frac{\lambda}{6.7 \times 10^{-12} \, \rm{cm}}\right)  \left(\frac{r_2}{10^6 \, \rm{cm}}\right) \, {\rm erg}.
\end{align}
If the flux tree is near the edge of the type-I region (i.e., $f_\rho \rightarrow 1$), one has 
\begin{align}
\label{eq:E_cfrho1}
    E_{\rm c} = \ &4.3 \times 10^{12} \times f_1^2 (1-f_\rho)^{3/2}   \left(\frac{H_{\rm c}}{10^{15}\, \rm{G}}\right)^2 \left(\frac{\delta_0}{-10^{-12} \, \rm{cm}}\right)  \nonumber \\ 
    & \times \left(\frac{\lambda}{6.7 \times 10^{-12} \, \rm{cm}}\right) \left(\frac{r_2}{10^6 \, \rm{cm}}\right) \, {\rm erg}.
\end{align}

\section{Magnetic field decay}
\label{sec:4}
The magnetic dipole moment, $\vec{\mu}$, can be inferred from the spin-down rate (magnitude of $\vec{\mu}$) and polarization swing (direction of $\vec{\mu}$) of a rotation-powered pulsar. In principle, it can be related to the magnetic field inside the star, although the relation is somewhat uncertain in the presence of a toroidal field component \citep{lasky_tilted_2013} and superconductivity \citep{lander_magnetic_2013}. 
The evolution of $\vec{\mu}$ and the internal field are linked. For example, the internal magnetic field decays, when the quantized flux tubes are transported from the outer core to the crust, where Ohmic dissipation occurs \citep{baym_electrical_1969}.
In Section \ref{sec:4.1}, we discuss pinning of quantized flux tubes to quantized vortices in the neutron superfluid (and vice-versa), and its implications for the magnetic field evolution.
We compare the vortex-flux-tube pinning strength, $E_{\rm pin}$, with $E_{\rm c}$ and find $E_{\rm c} \gg E_{\rm pin}$ under a range of plausible circumstances; that is, most quantized flux tubes are strongly coupled to their flux trees despite pinning.
In Section \ref{sec:4.2}, we discuss how flux coupling affects the macroscopic evolution of the magnetic field in type-I and -II superconductors. 
This analysis allows one to infer an upper limit for the size of the type-I superconducting region from a measurement of a neutron star's magnetic dipole moment.

\subsection{Neutron-vortex-proton-flux-tube pinning}
\label{sec:4.1}
In the outer core of a neutron star, the pinning of quantized proton flux tubes to the neutron superfluid vortices (and vice-versa) has been studied extensively in the literature \citep{muslimov_vortex_1985,sauls_superfluidity_1989,srinivasan_novel_1990,bhattacharya_evolution_1991,ruderman_neutron_1998,link_constraining_2003,glampedakis_magnetohydrodynamics_2011,glampedakis_magneto-rotational_2011,gugercinoglu_vortex_2014,alpar_flux-vortex_2017,drummond_stability_2017,drummond_stability_2018,thong_stability_2023}. 
Pinning hinders flux tubes from rising buoyantly into the resistive crust, lengthening the Ohmic decay time-scale \citep{baym_electrical_1969,srinivasan_novel_1990} of $\mu(t)$ \citep{srinivasan_novel_1990,ruderman_neutron_1998,ferrario_magnetic_2015}. Pinning is also cited as a mechanism for vortex avalanches (rotational glitches) and for linking rotational and magnetic evolution, e.g., $\mu(t) \propto \Omega(t)$, where $\Omega(t)$ is the magnitude of the angular velocity of the outer core \citep{srinivasan_novel_1990,bhattacharya_evolution_1991,ruderman_neutron_1998,alpar_flux-vortex_2017}. 
As electromagnetic and gravitational-wave torques decelerate the crust, vortices and flux tubes pinned to vortices are expelled from the outer core into the crust, thereby reducing $\mu(t)$ and $\Omega(t)$. 
\citet{srinivasan_novel_1990} invoked this mechanism to explain the contrast between the magnetic moments of isolated pulsars ($\mu \sim 10^{30} \, {\rm G \ cm^3})$ and recycled pulsars with an accretion history ($\mu \lesssim 10^{27} \, {\rm G \ cm^3}$).

\begin{figure*}
    \centering
    \includegraphics[width=1\textwidth]{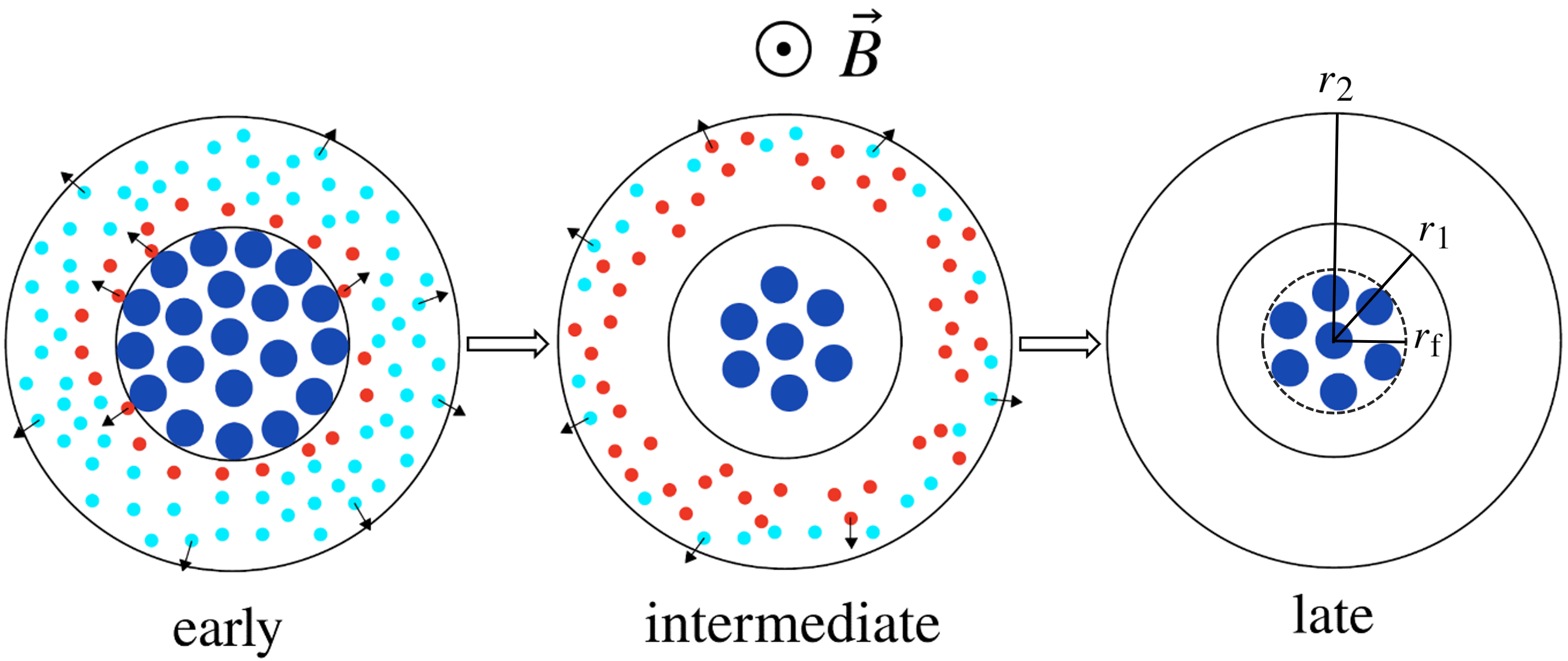}
    \caption{Schematic of neutron vortices dragging quantized flux tubes into the crust while leaving behind quantized flux tubes coupled to flux trees. The diagrams depict cross-sections through the equatorial plane of the inner and outer core, where the magnetic and rotation axes point out of the page. The boundary between the inner and outer core (outer core and crust) is represented by the inner (outer) black circle. The radii of the entire core, inner core and region containing frozen flux are denoted by $r_2, \, r_1,$ and $r_{\rm f}$, respectively. Blue-filled, large circles are macroscopic flux tubes in the inner core, together with quantized vortices (not drawn separately as their projections in the equatorial plane overlap) that are coupled strongly ($E_{\rm c} > E_{\rm pin}$) to flux trees. Cyan-filled, small circles are isolated quantized flux tubes, which are not part of a flux tree. Red-filled circles are weakly coupled ($E_{\rm c} < E_{\rm pin}$) quantized flux tubes unpinned by vortices from their flux trees. Small black arrows indicate the motion of quantized flux tubes dragged by vortices. The three diagrams are ordered chronologically from left to right, as the star spins down.}
    \label{fig:5}
\end{figure*}

How does $E_{\rm c}$ in Section \ref{sec:3} compare to the vortex-flux-tube pinning strength, $E_{\rm pin}$? 
The neutron superfluid and proton superconductor interact strongly via density and gradient (entrainment) coupling \citep{alpar_pinning_1977,sauls_superfluidity_1989,srinivasan_novel_1990,ruderman_neutron_1998}.
The dominant interaction between the topological defects of the two fluids is magnetic and arises from entrainment; neutron vortices entrain protons, dressing them with a magnetic flux $\sim 10^{-1} \Phi_0$, which pins them to the proton flux tubes with $1.6 \times 10^{-6} \lesssim |E_{\rm pin, max}|  / (1 \, {\rm erg}) \lesssim 8.0 \times 10^{-5} $ per vortex-flux-tube intersection, where $|E_{\rm pin, max}|$ is the magnitude of the maximum pinning energy, noting that the magnitude and sign of $E_{\rm pin}$ depend on the relative orientation of the vortex and quantized flux tube \citep{srinivasan_novel_1990,bhattacharya_evolution_1991,ruderman_neutron_1998,alford_flux_2008,alpar_flux-vortex_2017}.
We find $E_{\rm pin} \ll E_{\rm c}$ for most $f_1$ and $f_\rho$, except for $f_1 \ll 1$ as in equation (\ref{eq:E_cf10}), or $f_\rho \approx 1$ as in equation (\ref{eq:E_cfrho1}).
When a vortex pinned to a quantized flux tube moves outwards, it uncouples the flux tube from its flux tree for $E_{\rm pin} > E_{\rm c}$ by dragging it, or unpins from and cuts through the flux tube for $E_{\rm pin} < E_{\rm c}$.
Thus, vortices cut through quantized flux tubes in most situations, except when the type-I superconductor is small ($r_1 \ll r_2$, such that the macroscopic flux tube is shorter than the quantized flux tubes) or if the quantized flux tubes are coupled to a flux tree near the type-I-II interface ($\rho_{\rm t} \approx r_1$). 
Otherwise, vortices do not drag flux tubes away from their flux trees and into the crust.

\subsection{Frozen magnetic field and the maximum size of the type-I region}
\label{sec:4.2} 
Quantized flux tubes in the outer core of a neutron star fall into three categories: (i) isolated, i.e.\ they thread the type-II superconductor only and are not part of a flux tree; (ii) weakly coupled to a flux tree, i.e.\ $E_{\rm c} < E_{\rm pin}$; and (iii) strongly coupled to a flux tree, i.e.\ $E_{\rm c} > E_{\rm pin}$. Only categories (i) and (ii) can be pushed around by moving vortices; category (iii) is unaffected. 
Without a mechanism for moving an entire flux tree to the crust, where Ohmic decay occurs faster, the flux remains frozen within the inner core \citep{baym_electrical_1969,srinivasan_novel_1990}. 

The frozen magnetic field in the core may be the source of magnetic moments measured in recycled pulsars with a history of accretion from a binary companion.
Suppose first that all quantized flux tubes with $E_{\rm c}(f_1, f_\rho) < E_{\rm pin}$ are dragged out by vortices during the spin-down phase before accretion begins.
The remaining flux trees make up the residual magnetic field. 
Figure \ref{fig:5} presents an intentionally exaggerated schematic of the cross-sections through the equatorial plane of the inner and outer core to illustrate this process. 
Vortices drag isolated quantized flux tubes (depicted as small circles) into the crust. 
The flux tubes either do not belong to a flux tree (small, cyan-filled circles) or are weakly coupled to a flux tree before moving (small, red-filled circles). Strongly coupled flux tubes (not drawn in the cross-sections) are unaffected by the vortices, leaving behind frozen flux trees (large, blue-filled circles).
Consider frozen flux trees with $E_{\rm c} \geq E_{\rm pin}$ confined within the volume $r \leq r_{\rm f}$ in the type-I superconductor, while the strongly coupled quantized flux tubes occupy the volume $r_{\rm f} \leq r \leq r_2$ in the type-II superconductor.
The magnetic moment of this configuration amounts to 
\begin{equation}
\label{eq:mu_recyc}
    \mu_{\rm recyc} \approx B r_{\rm f}^2 r_2,
\end{equation}
where $\mu_{\rm recyc}$ is the magnetic moment of a recycled pulsar. 
In contrast, the magnetic moment of a newly born pulsar, $\mu_{\rm new}$, is given by
\begin{equation}
\label{eq:mu_new}
    \mu_{\rm new} \approx B r_2^3.
\end{equation}
Then, for $\mu_{\rm recyc} \lesssim 10^{27}  \, {\rm G\, cm^3}$ and $\mu_{\rm new } \sim 10^{30} \, {\rm G\, cm^3}$, equations (\ref{eq:mu_recyc}) and (\ref{eq:mu_new}) imply
\begin{align}
\label{eq:r_f}
    r_{\rm f} \sim 10^{4.5} \left(\frac{\mu_{\rm recyc}}{10^{27} \, {\rm G\, cm^3}}\right)^{1/2} \left(\frac{\mu_{\rm new}}{10^{30} \, {\rm G\, cm^3}}\right)^{-1/2} \left(\frac{r_2}{10^6 \, {\rm cm}}\right) {\rm cm}.
\end{align}

At $\rho_{\rm t} = r_{\rm f}$, we have $E_{\rm c} = E_{\rm pin}$; flux trees with $\rho_{\rm t} > r_{\rm f}$ are weakly coupled to their quantized flux tubes. Hence, we can substitute (\ref{eq:r_f}) into (\ref{eq:E_cfrho1}) to calculate $r_1>r_{\rm f}$.
We find 
\begin{align}
    r_1 - r_{\rm f} \approx & \ 5 \times 10^{-6} \left(\frac{r_{\rm f}}{10^6 \, \rm{cm}} \right)^{-1/3}\left(\frac{E_{\rm pin}}{8 \times 10^{-5} \, {\rm erg}}\right)^{2/3} \left(\frac{H_{\rm c}}{10^{15}\, \rm{G}}\right)^{-4/3}  \nonumber \\ 
    & \times \left(\frac{\delta_0}{-10^{-12} \, \rm{cm}}\right)^{-2/3} \left(\frac{\lambda}{6.7 \times 10^{-12} \, \rm{cm}}\right)^{-2/3}  \nonumber \\
    & \times \left(\frac{r_2}{10^6 \, \rm{cm}}\right)^{2/3} \, {\rm cm}.
\end{align}
That is, most non-isolated quantized flux tubes are strongly coupled to flux trees, and $r_1 \approx r_{\rm f}$.

\subsection{Idealisations}
\label{sec:4.3}
The above calculations regarding flux expulsion from the core are, of course, highly idealised.
In this section, we discuss some of the idealisations. 
Chief among them are the assumptions made about the macroscopic evolution and structure of the magnetic field.
We neglect expulsion of the magnetic field from the proton superconductor in the inner core by ambipolar diffusion \citep{passamonti_relevance_2017,kantor_note_2018}.
We also neglect the potentially significant toloidal component of the magnetic field \citep{braithwaite_axisymmetric_2009,lander_magnetic_2009,lander_magnetic_2013}. 
Toroidal flux branches in the same overall manner as poloidal flux if it threads the type-I and -II regions. 
However, the branching physics is more complicated.
The current-current vortex-flux-tube interaction \citep{ruderman_neutron_1998} depends on the relative orientation of an interacting vortex and flux tube, whereas the density interaction does not \citep{srinivasan_novel_1990}.
Additionally, the tangled poloidal and toroidal flux tubes may allow for easier expulsion of vortices through the multi-site unpinning mechanism termed ``slithering'' \citep{thong_stability_2023}; that is, more flux tubes may be ``left behind'' in the core, than if the magnetic field is purely poloidal.

A vortex can also pin to larger flux tubes near the type-I-II interface or in the type-I region.
For a flux tube with radius equal to $a_1/2^{M-1}$ to break off the trunk, where $M$ is some integer with $1 \leq M \leq N$, equation (\ref{eq:deltas}) takes the form 
\begin{equation}
    \Delta s =  2 \pi a_1 \left(1- \sqrt{1-4^{1-M}} -  2^{1-M} \right).
\end{equation}
Larger flux tubes (with smaller $M$) separating from the trunk have larger $\Delta s$, i.e.\ larger flux tubes require more energy to separate from their trunks.
A vortex can also pin to the trunk itself (largest flux tube in a flux tree), whereupon no flux tube separation needs to occur, for the flux tree to be expelled.
However, a vortex is unlikely to drag out an entire flux tree with $4^{N-1}-1$ quantized flux tubes, as the quantized flux tubes are pinned to other vortices in the outer core \citep{muslimov_vortex_1985,sauls_superfluidity_1989,srinivasan_novel_1990}. 
The situation is more complicated during a vortex avalanche, where many vortices migrate outwards coherently \citep{warszawski_gross-pitaevskii_2011}.
Under perfect conditions, it is conceivable that a flux tree can be dragged out in its entirety by a vortex avalanche for $\vec{\mu} \parallel \vec{\Omega}$, as all the vortices pull in the same direction. 
However, for $\vec{\mu} \nparallel \vec{\Omega}$, different vortices pinned to the same flux tree are likely to migrate towards the crust differently, e.g.\, vortices pinned in the northern hemisphere pull in the opposite direction to vortices pinned in the southern hemisphere.
The analysis of the collective interaction between vortex avalanches and flux trees for $\vec{\mu} \nparallel \vec{\Omega}$ is a challenging problem which should be studied with the help of numerical simulations and lies outside the scope of this paper.


\section{Conclusions}
In this paper, we calculate the free energy of macroscopic flux tubes branching into quantized flux tubes, as an idealised description of the magnetic coupling between the adjacent type-I and -II proton superconductors in the core of a neutron star. 
The study introduces a spatially varying normal-superconducting surface energy, $\gamma$, to model the type-I-II superconducting interface, extending previous studies where $\gamma$ is constant for a type-I superconductor in an externally applied and uniform magnetic field \citep{landau_intermediate_1938,landau_theory_1943}.
Assuming that $\gamma$ varies linearly with radius, we compare the free energy of superconductors with and without flux branching.
Without flux branching, $m$ flux tubes with radii $a_1/\sqrt{m}$ thread both superconductors rectilinearly.
We predict that flux branching is the preferred topology for a magnetic field threading both superconductors, for a wide range of plausible parameters.
Specifically, for all $m$ and $(L_1 + L_2)/2 \sim 10^6 \, {\rm cm}$, we find that flux branching minimizes the free energy for $8.6 \times 10^{-5} \lesssim L_2/L_1 \lesssim 1.4 \times 10^4$, $2.7 \times 10^{-6} \lesssim L_2/L_1 \lesssim 1.4 \times 10^2$, $8.4 \times 10^{-8} \lesssim L_2/L_1 \lesssim 0.87$ for $N = 28, 38, 48$, respectively. 
With flux branching, macroscopic flux tubes in the type-I superconductor divide dendritically into microscopic quantized flux tubes in the type-II superconductor. 
Near the type-I-II interface, flux tubes with varying numbers of integer $\Phi_0$ are found, similar to the type-II(n) flux tube structure obtained by \citet{alford_flux_2008} due to superfluid-superconductor coupling.

In Section \ref{sec:3}, we calculate $E_{\rm c}$, the energy required to separate a quantized flux tube from its flux tree.
The result is $0\lesssim E_{\rm c} / (1\, {\rm erg}) \lesssim 10^{13}$.
As a result, the magnetic coupling between most quantized and macroscopic flux tubes of the same flux tree greatly exceeds the strength of vortex-flux-tube pinning (i.e.\ $E_{\rm c} \gg E_{\rm pin})$, impeding magnetic decay and freezing flux trees in the core.
If this phenomenon causes the observed magnetic moments of recycled pulsars, one can infer the radius of the type-I superconductor within which frozen flux trees are found to be
\begin{align}
\label{eq:r_1_2}
    r_{1} \sim 10^{4.5} \left(\frac{\mu_{\rm recyc}}{10^{27} \, {\rm G\, cm^3}}\right)^{1/2} \left(\frac{\mu_{\rm new}}{10^{30} \, {\rm G\, cm^3}}\right)^{-1/2} \left(\frac{r_2}{10^6 \, {\rm cm}}\right) \, {\rm cm},
\end{align}
subject to the approximations discussed in Section \ref{sec:4.3} and elsewhere.

Strongly coupled ($E_{\rm c} \gg E_{\rm pin}$) quantized flux tubes are unaffected by vortices during spin down, resulting in $\mu(t) \not\propto \Omega(t)$.
That is, a neutron star can spin down without significant magnetic decay; only isolated and weakly coupled ($E_{\rm c} \ll E_{\rm pin}$) quantized flux tubes can be dragged by vortices to the crust for Ohmic decay.
In reality, of course, outwardly moving vortices may not remove all isolated and weakly coupled quantized flux tubes from the outer core during spin down.
If some isolated quantized flux tubes fail to be removed by vortices during spin down, the region $0 \leq r \leq r_{\rm f}$ defined by (\ref{eq:r_f}) contains isolated quantized flux tubes, and $r_1$ from (\ref{eq:r_1_2}) represents an upper limit.
There is a complex interplay between vortex and flux tube dynamics, influenced by factors such as interaction strength and tangled geometries \citep{drummond_stability_2017,drummond_stability_2018,thong_stability_2023}. 
Additionally, a neutron star may not contain a type-I superconductor at all, in which case the inference leading to (\ref{eq:r_1_2}) is void.
The inner core can contain other phases of dense matter, such as hyperonic matter \citep{vidana_hyperon-hyperon_2000,nishizaki_hyperon-mixed_2002,schaffner-bielich_phase_2002} and a quark color superconductor \citep{alford_magnetic_2000,alford_color-superconducting_2001,lattimer_physics_2004,alford_color-magnetic_2010,baym_hadrons_2018}. 
It is a question for future work, unanswered at the time of writing,  whether flux branching occurs between an adjacent type-I (or -II) proton superconductor and a color superconductor.

We emphasize that the treatment in this paper is highly idealized and relies on the following assumptions: (i) the normal-superconducting boundary is sharp, and flux-containing normal regions (flux tubes) do not interact \citep{chapman_asymptotic_1995}; (ii) $\gamma$ varies linearly with $r$, whereas in reality, the functional form of $\gamma$ depends on the density profile of the protons \citep{tinkham_introduction_2004}; (iii) the dynamical formation of flux trees is possible, which would require a time-dependent calculation; (iv) general relativity may be relevant to flux trees spanning the entire core; and 
(v) the observed magnetic moment does not trace the internal magnetic field exactly in the presence of toroidal components \citep{lander_magnetic_2013,lasky_tilted_2013,mastrano_neutron_2015} and diamagnetic screening from accreted matter \citep{romani_unified_1990,cumming_magnetic_2001,rai_choudhuri_diamagnetic_2002,payne_burial_2004}. 
Additionally, magnetic field lines with complicated geometries, like flux rings \citep{kozlov_closed_1993,cadorim_closed_2023} or tangled quantized flux tubes \citep{thong_stability_2023}, are neglected in this paper. 
These idealisations add to those discussed in Section \ref{sec:4.3}.

\section*{Acknowledgements}
The authors thank Yongyan Xia for helping to draw Figure \ref{fig:2}. 
We thank the anonymous referee for their constructive feedback and for supplying specific physical examples to bolster the discussion in Section \ref{sec:4.3}.
Parts of this research are supported by an Australian Government Research Training Program Scholarship (Stipend), Research Training Program Scholarship (Fee Offset), Rowden White Scholarship, McKellar Prize in Theoretical Physics, Professor Kernot Research Scholarship in Physics and the Australian Research Council (ARC) Centre of Excellence for Gravitational Wave Discovery (OzGrav) (grant number CE170100004).

\section*{Data Availability}
No new data were generated or analysed in support of this research.



\bibliographystyle{mnras}
\bibliography{references} 




\appendix

\section{Type-I and -II Superconductivity}
\label{appendix:A}
Type-I superconductors have $\delta > 0$, while type-II superconductors have $\delta < 0$. 
Consequently, type-I minimizes and type-II maximizes $S$ from (\ref{eq:F1}). 
The competition governs the size and number of the normal regions in type-I and -II superconductors. 
Consider a superconductor with magnetic flux, $\Phi$, in the following two scenarios: 
\begin{enumerate}
    \item $\Phi$ is entirely contained within a macroscopic cylindrical normal region with radius $a_1$; and
    \item $\Phi$ is uniformly distributed over $n$ smaller cylindrical regions with radii $a_n < a_1$, with $n > 1$. 
\end{enumerate}
Flux conservation means $a_1 = \sqrt{n} a_n$. Hence, the surface energy per unit length is given by $2\pi a_1$ and $n^{-1/2} 2 \pi a_1$ for (i) and (ii), respectively. Scenario (i) with a macroscopic normal region is energetically preferred for $\delta > 0$ (type-I),  while scenario (ii) with many smaller normal regions is energetically preferred for $\delta < 0$ (type-II). 

\section{Branch opening angle}
\label{appendix:B}
In this appendix, we discuss how wide the flux trees are in a neutron star, i.e.\ how closely packed they are in the type-I superconducting region. 
The aim is to clarify the assumptions underpinning (\ref{eq:d1}) and (\ref{eq:Nft}) and hence the calculation of the surplus free energy $F_{\rm b}$. 

Among other things, the width of a flux tree is set by $d_1$ or equivalently the branch half opening angle $\theta_1$ in Figure \ref{fig:1}. 
Upon inspecting Figure \ref{fig:3} and equations (\ref{eq:a_n}) and (\ref{eq:d_n}), we see that the width of a flux tree equals 
\begin{equation}
    \sqrt{2}\sum_{n=1}^{N-1} \left(d_n + a_{n+1}\right) = 2\sqrt{2} \left(d_1+a_2\right)
\end{equation}
and $d_1$ determines the characteristic
separation, $\sqrt{2}\left(d_{N-1}+a_{N}\right)$, between the quantized flux tubes (smallest branches) which terminate the tree in the type-II superconductor in the outer core. 
Equation (\ref{eq:FNI_i}) and Figure \ref{fig:3} imply that $d_1 = (\sqrt{2}-1)a_1/2$ minimizes (\ref{eq:FNI_i}) without the normal regions overlapping.

The sharp-interface approximation assumes that normal regions do not interact.
In reality, quantized flux tubes in the type-II superconductor repel strongly when separated by $\sim \lambda$.
Strictly speaking, $d_1$ is determined by the competition between flux-tube-flux-tube repulsion and energy minimization of the flux tree.
We will address this challenging problem in a forthcoming paper. 
For now, we assume provisionally that the flux trees are as wide as they can be without overlapping, maximizing $d_1$, so that the quantized flux tubes interact as weakly as possible.
With this assumption, flux trees fill the type-I superconductor without gaps, implfying
\begin{equation}
    d_1 = \sqrt{\frac{\pi (L_1/2)^2 }{8 N_{\rm ft}}} - a_2,
\end{equation}
where $N_{\rm ft}$ is the number of flux trees implied by flux conservation, given by
\begin{equation}
    N_{\rm ft} = \frac{B (L_1/2)^2}{H_{\rm c} a_1^2}.
\end{equation}
In a neutron star, one has $B/H_{\rm c} \sim 10^{-3}$ and hence $d_1 \approx 19 a_1$, which we assume throughout this paper. 
The latter value of $d_1$ is approximately two orders of magnitude larger than the minimum value $d_1 = (\sqrt{2}-1)a_1/2$.

\bsp	
\label{lastpage}
\end{document}